\documentclass[fp,twocolumn]{jpsj3}
\usepackage{txfonts}
\usepackage{braket}
\usepackage{amsmath,amssymb}
\usepackage{graphicx,xcolor}
\usepackage{here}
\usepackage{color}
\usepackage[hyphens]{url}

\title{Impact of Fixing Spins in a Quantum Annealer with Energy Rescaling}

\author{Tomohiro Hattori$^1$, Hirotaka Irie$^2$, Tadashi Kadowaki$^{2,3}$ and Shu Tanaka$^{1,4,5,6}$\thanks{shu.tanaka@appi.keio.ac.jp}}
\inst{$^1$ Graduate School of Science and Technology, Keio University, 3-14-1 Hiyoshi, Kohoku-ku, Yokohama-shi, Kanagawa 223-8522, Japan \\
$^2$ DENSO CORPORATION, Haneda Innovation City D3, 1-1-4, Hanedakuko, Ota-ku, Tokyo 144-0041, Japan \\
$^3$ National Institute of Advanced Industrial Science and Technology (AIST), 1-1-1, Umezono, Tsukuba-shi, Ibaraki 305-8568, Japan\\
$^4$ Department of Applied Physics and Physico-Informatics, Keio University, 3-14-1 Hiyoshi, Kohoku-ku, Yokohama-shi, Kanagawa 223-8522, Japan \\
$^5$ Keio University Sustainable Quantum Artificial Intelligence Center (KSQAIC), Keio University, Tokyo 108-8345, Japan \\
$^6$ Human Biology-Microbiome-Quantum Research Center (WPI-Bio2Q), Keio University, Tokyo 108-8345, Japan
}
\abst{
Quantum annealing is a promising algorithm for solving combinatorial optimization problems.
However, various hardware restrictions significantly impede its efficient performance.
Size-reduction methods provide an effective approach to addressing large-scale problems but often introduce additional difficulties.
A notable hardware restriction is that quantum annealing can handle only a limited number of decision variables, compared to the size of the problem.
Moreover, when employing size-reduction methods, the interactions and local magnetic fields in the Ising model––used to represent the combinatorial optimization problem––can become excessively large, making them difficult to implement on hardware.
Although prior studies suggest that energy rescaling impacts the performance of quantum annealing, its interplay with size-reduction methods remains unexplored.
This study examines the relationship between fixing spins, a promising size-reduction method, and the effects of energy rescaling.
Numerical simulations and experiments conducted on a quantum annealer demonstrate that the fixing spins method enhances quantum annealing performance while preserving the spin-chain embedding for a homogeneous, fully connected ferromagnetic Ising model.
}

\begin{document}
\maketitle

\section{\label{introduction}Introduction}

Combinatorial optimization problems are ubiquitous across various domains, necessitating high-accuracy solutions to address their complexity. 
A critical challenge in these problems is the exponential growth of solution candidates as the problem size increases.
With the ongoing advancement of information-driven society, the demand for solving large-scale optimization problems has surged. 
Quantum annealing~\cite{Kadowaki1998Quantum, FINNILA1994343QA, Das2008qa, application_software, QAreview2023} has emerged as a promising metaheuristic to effectively tackle these challenges.

Recently, quantum annealing has been applied across diverse fields, including logistics~\cite{Weinberg2023logistics}, materials science~\cite{camino2023quantum, honda2024development, xu2025quantum}, and finance~\cite{rosenberg2015solving_trading, grant2021benchmarking_portofolio}. 
Simultaneously, significant efforts have been devoted to developing and refining algorithms that leverage the potential of quantum annealing.
This growing body of research underscores the importance of advancing quantum annealing methodologies for various combinatorial optimization applications. 

Quantum annealing operates by searching for the ground state of the Ising model, which can be mapped to the solutions of combinatorial optimization problems. 
While many such problems can be formulated as Ising models, certain types, including those involving integer variables or black-box optimization, cannot be directly expressed in this framework. 
However, recent advancements are rapidly expanding the scope of quantum annealing, enabling it to address these complex problems~\cite{FMA2020, koshikawa2021benchmark, FMA2022, matsumori2022application, FMA2023, photonic_laser2022, xu2025quantum}.
Furthermore, hybrid approaches that incorporate quantum annealers into intermediate stages of algorithms are broadening their applicability~\cite{hirama2023efficient, kanai2024annealing}.

Despite these advancements, significant challenges remain in optimizing hardware for quantum annealing.
These challenges can be categorized into three primary constraints: size restrictions, graph structure restrictions, and energy-scale restrictions.

Size restrictions are a critical barrier, limiting quantum annealers' ability to solve large-scale problems.
Presently, the size of solvable problems is restricted by the number of qubits available on quantum annealing devices. 
As problem sizes grow, maintaining solution quality necessitates considerably longer annealing times~\cite{altshuler2010anderson, hattori2025advantages}.
Additionally, decoherence effects increasingly hinder the performance of quantum annealers over extended durations, further complicating efforts to address large-scale optimization problems effectively.

Various methods have been proposed to mitigate the hardware size restrictions of quantum annealers, particularly through size-reduction techniques aimed at solving large-scale problems. 
One study introduced an approach to efficiently search for solutions to large-scale combinatorial optimization problems by classifying spins based on their convergence behavior in molecular dynamics simulations~\cite{Irie2021Hybrid}.
Spins exhibiting slow convergence are deemed challenging to resolve and are delegated to quantum annealing for determination. 

Other methods leverage sample persistence~\cite{Karmi2017, Karimi2017_2, Atobe_2022, kikuchi2023hybrid}, which originates from the concept of persistency~\cite{Hammer_1984}.
Persistency involves rigorously solving specific quadratic pseudo-functions.
However, solving a broad range of combinatorial optimization problems remains difficult.
Classical methods assume that spins maintaining the same orientation after multiple iterations are likely to persist.
These strategies enable quantum annealers to address large-scale problems and produce high-quality solutions in practice. 
Nevertheless, the precise methodologies for resizing problems lack clarity, underscoring the importance of investigating the effects of problem size reduction.

In addition to size restrictions, the hardware limitations associated with graph structures in quantum annealers contribute to suboptimal solutions. 
Ising models with denser graphs than those supported by the quantum annealer's architecture cannot be faithfully represented. 
To bridge this gap, embedding methods have been developed~\cite{Choi_2008, Choi_2011}.
These techniques utilize qubit copies, referred to as {\it qubit chains}, to effectively represent the couplers of dense Ising models. 
While such embeddings allow quantum annealers to handle dense graphs, they require a significant number of qubits.
Here, the length of the qubit copies is called {\it chain length}.
The qubit chains and demand careful parameter tuning, particularly the {\it chain strength} between qubit copies. Setting optimal parameters is a challenging task, making embedding one of the factors that can compromise solution quality.

Several approaches have been proposed to alleviate the graph structure restrictions in quantum annealers. 
The first approach focuses on reducing the number of spins required for embedding by increasing the graph density on the hardware. 
This effort has led to the development of increasingly dense graph architectures~\cite{dwave_prototype}.
% Additionally, the LHZ scheme~\cite{LHZ2015quantum} is widely recognized as an efficient method for implementing fully connected graphs on practical quantum annealing devices. 

The second approach seeks to minimize the embedding overhead by reducing the connectivity between qubits by reducing the problem size.
In dense graph scenarios, the number of required couplings grows significantly with the number of qubits. 
For instance, a fully connected graph with $n$ spins requires $n-1$ couplings, demanding graph structures with higher connectivity as the number of qubits increases.

The third approach involves error mitigation techniques. 
Previous studies have explored methods to mitigate errors and maintain solution accuracy by parallelizing the graph structure~\cite{vinci2015quantumerror, vinci2016nested, Pelofske_2022, hino2024physical}.
Collectively, these approaches demonstrate diverse strategies to address the graph structure constraints of quantum annealers.
Rescaling techniques are also employed to embed the Ising model onto a practical quantum annealer, overcoming limitations related to energy scales. 
Recent studies suggest that energy rescaling impacts noise strength~\cite{amin2023quantummitigation}.
Leveraging this property, ideal energy parameters have been estimated using zero-noise extrapolation. 
Another study treated rescaling parameters as variational parameters, showing that optimized rescaling significantly enhances performance~\cite{Braida2024rescalingparameter}. 
Therefore, incorporating rescaling considerations is essential for achieving high-quality solutions.

In this study, we explore the interplay between size-reduction techniques and rescaling techniques. 
Implementing size-reduction methods modifies the Ising model's properties, often resulting in large local magnetic field values. This alteration is closely linked to energy rescaling, highlighting the need to investigate their relationship comprehensively.

The remainder of this paper is structured as follows. Section~\ref{setup} details the variable fixing and rescaling methods used in this study and outlines the problem setup. 
Section~\ref{results} presents the results for the homogeneous, fully connected ferromagnetic Ising model. 
Finally, Section~\ref{conclusion} summarizes the findings and discusses potential future directions.

\section{\label{setup}Setup}
\begin{figure*}[t]
    \centering
    \includegraphics[clip,scale=0.5]{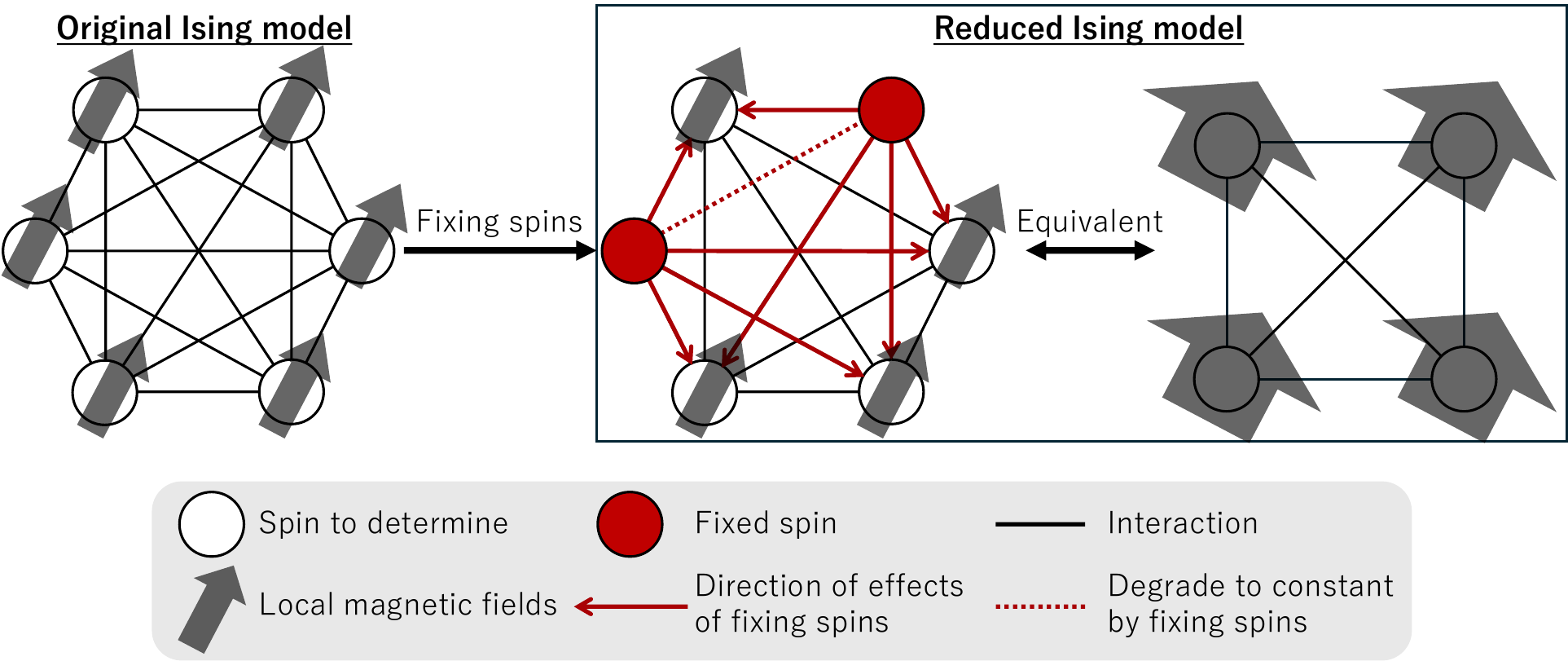}
    \caption{(Color online) Conceptual diagram illustrating the fixing spins method. This method reduces the size of the Ising model while altering the properties of local magnetic fields. The graph on the left represents the original Ising model, where white circles indicate spins determined by quantum annealing, and red circles represent fixed spins. Black arrows denote the local magnetic fields at each vertex, while solid lines indicate the interactions of the Ising model. The dotted line represents an interaction that simplifies to a constant due to the fixed spins. Dark red arrows illustrate the directional effects of the fixed spins on the system. The graph on the right depicts the reduced Ising model, where the influence of the fixed spins enhances the local magnetic fields of the simplified problem. In this instance, a fully connected Ising model with $N=6$ is reduced to a fully connected Ising model with $N=4$.}
    \label{fig:fixing_spin}
\end{figure*}

In quantum annealing, solutions to a combinatorial optimization problem are obtained by mapping them to the ground state of the Ising model.
The Hamiltonian of the Ising model, representing the problem to be solved, is denoted as $\mathcal{H}_\mathrm{p}$.
Quantum fluctuations, which facilitate the exploration of the solution space, are represented by the driver Hamiltonian $\mathcal{H}_\mathrm{d}$. 
The total Hamiltonian governing the quantum annealing process is expressed as:
\begin{align}
    \label{eq:H}
    \mathcal{H}(s) = A(s)\mathcal{H}_\mathrm{d} + B(s)\mathcal{H}_\mathrm{p},\quad 0\leq s\leq 1.
\end{align}
Here, $ s \in [0,1] $ denotes the normalized time, where the annealing time $\tau$ relates to the normalized time as $s =\frac{t}{\tau}$, with $t$ being the elapsed time.
The annealing schedules $A(s)$ and $B(s)$ are monotonically decreasing and increasing functions, respectively, satisfying $A(0)\gg B(0)$ and $A(1)\ll B(1)$.

The quantum state is expected to adiabatically follow the ground state of the time-dependent Hamiltonian $\mathcal{H}(t) $, ultimately reaching the ground state of the Ising model at the final annealing time.
According to the adiabatic theorem~\cite{Messiah1976adiabatictheorem}, the annealing time must be sufficiently long to ensure adiabatic evolution, which is determined by the minimum energy gap $\Delta_\mathrm{min}$ between the ground state and the first excited state during the process. 
Consequently, $\Delta_\mathrm{min}$ serves as a critical performance metric for quantum annealing.

In this study, transverse local magnetic fields are employed as the driver Hamiltonian as $\mathcal{H}_\mathrm{d}$. 
The problem Hamiltonian $\mathcal{H}_\mathrm{p}$ and the driver Hamiltonian $\mathcal{H}_\mathrm{d}$ are expressed as follows:
\begin{align}
    \label{eq:Hp}
    \mathcal{H}_\mathrm{p} &= -\sum_{i=1}^{N}h_i\sigma_i^z - \sum_{1 \leq i < j \leq N}J_{ij}\sigma_i^z\sigma_j^z,\\
    \label{eq:Hd}
    \mathcal{H}_\mathrm{d} &= -\sum_{i=1}^{N}\sigma_i^x,
\end{align}
where $N$ represents the number of spins, $h_i$ represents the local magnetic field acting on the $i$th spin, $J_{ij}$ represents the interaction between the $i$th and $j$th spins, and $\sigma_i^x, \sigma_i^z$ are Pauli matrices associated with the $i$-th spin component.

Upon applying the fixing spins method, the Hamiltonians are modified. 
The following explains the fixing spin method, in which some spin variables are fixed at $+1$ or $-1$.
The updated problem Hamiltonian, $\mathcal{H}_\mathrm{p}^{\prime}$, and the driver Hamiltonian, $\mathcal{H}_\mathrm{d}^{\prime}$, after fixing spins are given as follows:
\begin{align}
    \label{eq:Hp_fixed}
    \mathcal{H}_\mathrm{p}^{\prime} &= -\sum_{i^\prime=1}^{n}h_{i^\prime}^{\prime}\sigma_{i^\prime}^z - \sum_{1 \leq i^\prime < j^\prime \leq n}J_{i^\prime j^\prime}^{\prime}\sigma_{i^\prime}^z\sigma_{j^\prime}^z,\\
    \label{eq:Hd_fixed}
    \mathcal{H}_\mathrm{d}^{\prime} &= -\sum_{i^\prime=1}^{n}\sigma_{i^\prime}^x,
\end{align}
where $n$ denotes the number of spins after applying the fixing spins method, and $i^{\prime}, j^{\prime}~(=1,\dots,n)$ are the indices corresponding to the spins after fixing spins.
The parameters $h_{i}^{\prime}$ and $J_{ij}^{\prime}$ are expressed as follows:
\begin{align}
    \label{eq:h_fixed}
    h_{i^\prime}^{\prime} &= h_{i^\prime} + \sum_{k^\prime=n+1}^N J_{i^\prime k^\prime}s_{k^\prime},\\
    \label{eq:J_fixed}
    J_{i^\prime j^\prime}^{\prime} &= J_{i^\prime j^\prime},
\end{align}
where $s_k = \pm 1$ represents the fixed spin.
As shown in Eq.~\eqref{eq:h_fixed} and Eq.~\eqref{eq:J_fixed}, the interaction terms involving fixed spins reduce to local magnetic fields, while the interaction terms not connected to the fixed spins remain unchanged.
Figure \ref{fig:fixing_spin} illustrates the conceptual diagram of fixing spins in a fully connected Ising model with $N=6$.
The local magnetic field terms at fixed spins and the interaction terms between fixed spins become constants, meaning they do not contribute to the reduced problem.
Additionally, the interaction terms between a fixed spin and a spin in the reduced problem transform into local magnetic fields. 
Therefore, the effects of fixing spins are twofold: reducing the number of spins and altering the properties of the local magnetic fields.

The selection of spins to be fixed, along with their respective directions, is determined using a classical pre-processing method from previous studies~\cite{Irie2021Hybrid, Karmi2017, Karimi2017_2, Atobe_2022, kikuchi2023hybrid}. 
In this study, to characterize the fixed spins without assuming a predetermined selection, we adopt the error probability $ p_\mathrm{err} $ as defined in the previous study~\cite{hattori2025advantages}.
The error probability $ p_\mathrm{err} $ is the proportion of spins that are aligned with the opposite direction of the ground state of the Ising model within the spin configuration.
The previous study shows the effectiveness of fixing spins for large-scale combinatorial optimization problems in the quantum annealer.
Also, the fixing spin method enhances the performance of the quantum annealing from the perspective of the energy gap.
However, the impact of the error of fixing spins on the energy gap is not revealed.
In this study, the error probability is easy to calculate since a homogeneous fully connected ferromagnetic Ising model, which has the trivial ground state, is used.

The input parameter ranges for hardware are defined as $[\mathcal{M}_\mathrm{min}, \mathcal{M}_\mathrm{max}]$ for the local magnetic field and $[\mathcal{J}_\mathrm{min}, \mathcal{J}_\mathrm{max}]$ for the interaction.
These parameters $h_i$ and $J_{ij}$ are rescaled to fit within these ranges by dividing by the rescaling parameter $r$.
In the previous study~\cite{hattori2025advantages}, the rescaling parameter $r$ is defined as follows:
\begin{align}
    \label{eq:rescaling_parameter}
    r = \max\left\{\frac{\max\{h_{i}\}}{\mathcal{M}_\mathrm{max}},\frac{\min\{h_{i}\}}{\mathcal{M}_\mathrm{min}}, \nonumber\right.\\\left. \frac{\max\{J_{ij}\}}{\mathcal{J}_\mathrm{max}}, \frac{\min\{J_{ij}\}}{\mathcal{J}_\mathrm{min}}\right\}.
\end{align}
In this study, the primary objective is to explore the effects of rescaling. 
To this end, we investigated the dependence of quantum annealing performance on the rescaling parameter $r$.
The Hamiltonian after rescaling is expressed as follows:
\begin{align}
    \label{eq:rescaling_Hp}
    \mathcal{H}_\mathrm{p}^{\prime\prime} = \frac{\mathcal{H}_\mathrm{p}^{\prime}}{r}.
\end{align}
The quantum annealing Hamiltonian after rescaling is expressed as follows:
\begin{align}
    \label{eq:rescaling_H}
    \mathcal{H}^\prime(t) = A(s)\mathcal{H}_\mathrm{d}^\prime + B(s)\mathcal{H}_\mathrm{p}^{\prime\prime}.
\end{align}

The rescaling parameters are affected by fixing spins because the local magnetic fields are changed as shown in Eq.~\eqref{eq:h_fixed}.
Consequently, the fixing spins method alters the range of local magnetic fields, thereby influencing the rescaling parameters.
This study examines the relationship between the fixing spins method and rescaling.
A previous study~\cite{hattori2025advantages} demonstrated performance improvements through fixing spins.
However, this conclusion may not hold when rescaling is considered.

\section{\label{results}Results}
\begin{figure*}[t]
    \begin{tabular}{ccc}
    \begin{minipage}[c]{0.33\hsize}
    \includegraphics[clip,scale=0.9]{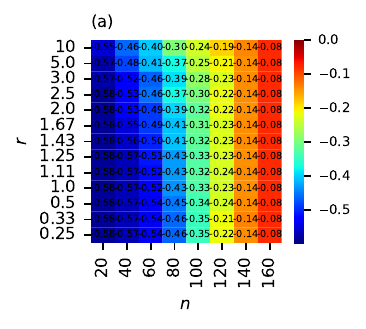}
    \end{minipage}
    \begin{minipage}[c]{0.33\hsize}
    \includegraphics[clip,scale=0.9]{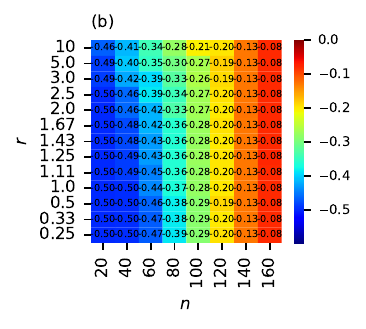}
    \end{minipage}
    \begin{minipage}[c]{0.33\hsize}
    \includegraphics[clip,scale=0.9]{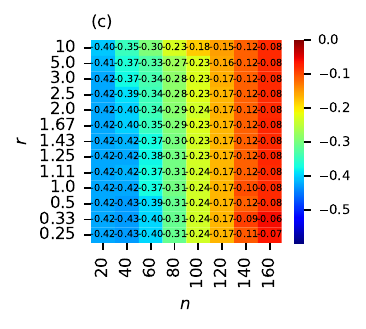}
    \end{minipage}\\
    \begin{minipage}[c]{0.33\hsize}
    \includegraphics[clip,scale=0.9]{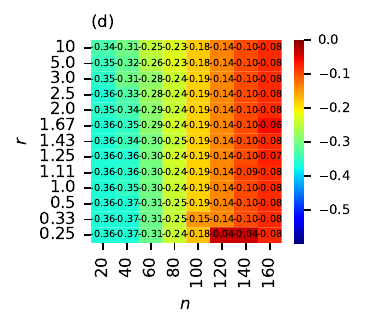}
    \end{minipage}
    \begin{minipage}[c]{0.33\hsize}
    \includegraphics[clip,scale=0.9]{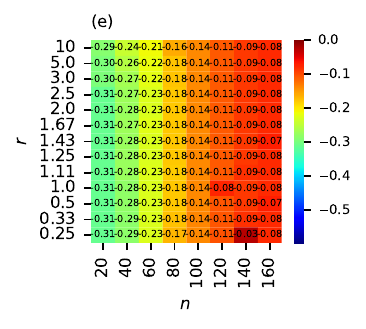}
    \end{minipage}
    \begin{minipage}[c]{0.33\hsize}
    \includegraphics[clip,scale=0.9]{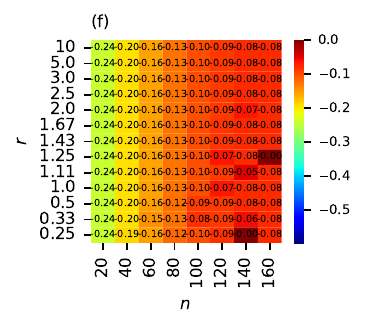}
    \end{minipage}
    \end{tabular}
    \caption{(Color online) The color map shows the distribution of the minimum energy in the $n$ versus $r$ space for the homogeneous fully connected ferromagnetic Ising model with $N=160$. The spacing of $r$ values is not necessarily uniform. $n=160$ corresponds to the results without fixed spins. The annealing time is set to $\tau=2000.0~\mu s$. The ground state energy is $0.597$. Each panel represents a different error rate for fixing spins, with $p_\mathrm{err}$ set to (a) $p_\mathrm{err}=0.0$, (b) $p_\mathrm{err}=0.1$, (c) $p_\mathrm{err}=0.2$, (d) $p_\mathrm{err}=0.3$, (e) $p_\mathrm{err}=0.4$ and (f) $p_\mathrm{err}=0.5$, respectively.}
    \label{fig:error_res_n_min_energy}
\end{figure*}
In this study, for simplicity, we focus on a homogeneous fully connected ferromagnetic Ising model, as the Ising model obtained after fixing spins results in another fully connected Ising model.
Thus, the model generated after fixing spins will not include isolated spins, making it easier to analyze.

The homogeneous fully connected ferromagnetic Ising model is expressed as follows:
\begin{align}
    \label{eq:ferr_Hp}
    \mathcal{H}_\mathrm{p}=-h\sum_{i=1}^N\sigma_i^z-J/N\sum_{1\leq i<j\leq N}\sigma_i^z\sigma_j^z.
\end{align}
The normalization by $1/N$ ensures that $\mathcal{H}_\mathrm{p}$ is an extensive quantity.
If the ground state is degenerate, determining the error rate becomes difficult because the ground state is not uniquely defined.
Therefore, we consider the Ising model without degeneracy.
The ferromagnetic Ising model with $h=0$ exhibits trivial twofold degeneracy.
To eliminate this trivial degeneracy, we set $h=0.1$ and $J=1$.
The ground state of this homogeneous fully connected Ising model is trivially the all-up state.

The D-Wave Advantage system 6.4~\cite{D-Wave_Advantage} was used as the quantum annealer for the experiment. 
This system features $5760$ physical qubits arranged in the Pegasus graph~\cite{D-Wave_Topologies}, ideally enabling the embedding of a complete graph with $177$ nodes through minor embedding~\cite{Choi_2008, Choi_2011}.
The chain length is determined by the graph topology of the qubits on hardware, depending on the number of couplers of each qubit.
However, not all qubits in the system are functional.
Consequently, the number of spins in the Ising model was set to $N=160$, the near maximum size that can be embedded in the quantum annealer. 

The Ising model is embedded in the quantum annealer using minor embedding, and the chain strength—representing the interaction between physical qubits—must be adjusted to ensure that qubits within the same qubit chain align in the same direction.
When we naively solve the problems in quantum annealer, the {\it auto-scale} method is employed~\cite{Ocean}.
In the algorithm of {\it auto-scale} method, the longer length of the qubit chain requires larger values of chain strength.
However, the chain strength must remain within the range $[\mathcal{J} _\mathrm{min},\mathcal{J}_\mathrm{max}]$ for the interaction.
For this study, the chain strength was set to the fixed value of $2.0$, the strongest value permissible in this configuration\cite{Ocean}. 
Therefore, the relative chain strength varies depending on the parameter range of the Ising model.
When the parameter range of the Ising model is large, the chain strength becomes relatively small.
Due to the stochastic behaviour, the ground state search in the quantum annealer was performed $100$ times, and the minimum energy from these runs was adopted as the quantum annealer's energy output in this experiment.

Additionally, we investigated the energy gap using exact diagonalization for the same fully connected Ising model without considering the embedding effect.
The energy gap analysis yielded similar results for the long annealing time limit.
Finally, we investigated the system's properties at the thermodynamic limit, accounting for the effects of rescaling and fixing spins.

The annealing schedule after rescaling is shown in appendix~\ref{appendix_annealing_sche}.

\subsection{Experiment on an actual quantum annealer}
When embedding an Ising model with a large energy scale into a real quantum annealer, rescaling the energy scale becomes unavoidable.
This rescaling is performed as described in Eq.~\eqref{eq:rescaling_parameter}.
The experiment was conducted on the D-Wave Advantage system 6.4 to examine how the quantum annealer's performance depends on the rescaling parameter.

Figure~\ref{fig:error_res_n_min_energy} illustrates the parameter dependence of the minimum energy among $100$ runs.
% Since the energy scale after rescaling differs from the pre-rescaling energy scale, direct comparison of results is challenging.
% Therefore, the energy values are restored to the original Ising model, as expressed in Eq.~\eqref{eq:ferr_Hp}, after solving them with the quantum annealer.
Since the energy scale after rescaling differs from the pre-rescaling energy scale, the energy values are restored to the original Ising model, as expressed in Eq.~\eqref{eq:ferr_Hp}, after solving them with the quantum annealer.
The $x$-axis represents the number $n$ of spins after fixing spins, and the $y$-axis denotes the rescaling parameter $r$, as shown in Eq.~\eqref{eq:rescaling_Hp}.
As $r$ decreases, the quantum annealer achieves lower energy, as shown in Fig~\ref{fig:error_res_n_min_energy}~(a)-(c).
This result indicates that a wide parameter range is necessary to maintain the performance of the quantum annealer.

Conversely, the dependence on $r$ gradually diminishes in both high error probability scenarios and when the size of the reduced problem increases, as seen in Fig.~\ref{fig:error_res_n_min_energy}.
The behavior of the local magnetic fields in the reduced problems after fixing spins is consistent across these scenarios.
Errors in fixing spins affect the local magnetic fields of the reduced problem.
The upper bound of the absolute value of the local magnetic fields after fixing spins is given by:
\begin{align}
    \label{eq:h_value_increase}
    h^{\prime}_i = h_i + \sum_{k=n+1}^N J_{ik} s_k \leq h_i + \sum_{k=n+1}^N |J_{ik}|.
\end{align}
Equation~\eqref{eq:h_value_increase} exhibits the maximum increase in the range of the local magnetic field due to fixed spins.
Since we are dealing with a homogeneous fully connected ferromagnetic Ising model, after fixing spins with $p_\mathrm{err}=0$, the local magnetic fields correspond to this upper bound.
When fixing spins with $p_\mathrm{err}=0$, the local magnetic fields change as follows:
\begin{align}
    \label{eq:h_analysis}
    h^{\prime} = h + J(N-n).
\end{align}
Following Eq.~\eqref{eq:h_fixed}, the homogeneous fully connected ferromagnetic Ising model after fixing spins with $p_\mathrm{err}$ is expressed as follows:
\begin{align}
    \label{eq:h_analysis_error}
    h^{\prime} = h + J(N-n)(1-2p_\mathrm{err}).
\end{align}
Equation~\eqref{eq:h_analysis_error} indicates that $h^{\prime}$ decreases as $p_\mathrm{err}$ and $n$ increase.
Consequently, the effects of rescaling become more pronounced when $p_\mathrm{err} = 0$.
According to Eq.~\eqref{eq:h_analysis_error}, $h^{\prime}$ equals $h$ when $p_\mathrm{err}= 0.5$ or $n=N$.
In this case, the energy range of the Ising model remains unchanged.
Therefore, fixing spins does not affect the energy range, and the dependence on $r$ disappears.

The parameters associated with qubit chains are crucial for performing QA to the Ising model with dense interactions.
Ideally, each qubit chain should correspond to a single logical spin, with all qubits within the chain aligned in the same direction.
To ensure this, the chain strength should be set as large as possible.
In this experiment, we fix the chain strength at its maximum value.
However, even at maximum chain strength, qubits within a qubit chain may still fail to align uniformly, leading to broken qubit chains.
High energy points observed in Fig.~\ref{fig:error_res_n_min_energy} (d)-(f) result from the breaking of the embedding chain in the hardware.

When the chain strength is fixed, the ratio of the local magnetic fields to the chain strength becomes a critical factor.
The local magnetic field is distributed equally among the qubits within the qubit chain.
Assigning sufficiently large values of the local magnetic fields is also an effective strategy for preventing the qubit chain from breaking.
Therefore, the longer the chain length, the weaker local magnetic fields are assigned to each qubit.
Consequently, as the spin length increases, the ratio of the local magnetic fields to the chain strength decreases.
Therefore, with a small $h$, maintaining the qubit chain becomes increasingly difficult as its length grows.

When $n$ is large, the long qubit chain is required to express the Ising model in hardware.
As a result, large $n$ leads to a small ratio of the local magnetic fields to the chain strength.
In addition, increasing both $n$ and $p_\mathrm{err}$ results in a smaller $h^\prime$ from Eq.~\eqref{eq:h_analysis_error}.
Consequently, with large $n$ and $p_\mathrm{err}$, the qubit chain is more likely to break compared to other parameter ranges.

When $r$ is small, the ratio of the local magnetic fields to the chain strength becomes small, which renders the qubit chain susceptible to chain breaking.
To set the larger ratio of the local magnetic fields to the chain strength is necessary to avoid chain break for small $r$.
Therefore, expanding the parameter range is necessary to accommodate large-scale problems when embedding into the quantum annealer and preventing chain break.
These results highlight the importance of fixing spins in the correct direction to achieve lower energy and preserve qubit chains on the hardware.
\begin{figure}[t]
    \centering
    \includegraphics[clip,scale=1.0]{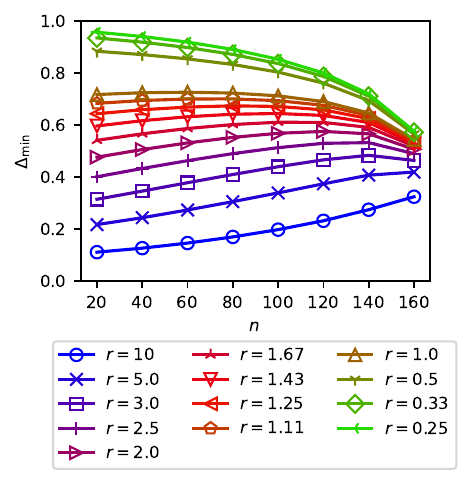}
    \caption{(Color online) The minimum energy gap $\Delta_\mathrm{min}$ of the Hamiltonian after fixing spins with $p_\mathrm{err}=0$ and rescaling with $r$ for the homogeneous fully connected ferromagnetic Ising model with $N=160$. $n=160$ corresponds to the results without fixing spins. Solid lines between points are provided as a guide to the eye.}
    \label{fig:eg_rescaled}
\end{figure}
\begin{figure*}[t]
    \begin{tabular}{ccc}
    \begin{minipage}[c]{0.33\hsize}
    \includegraphics[clip,scale=0.9]{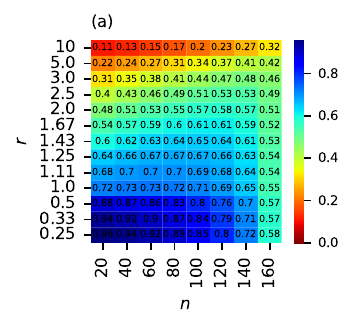}
    \end{minipage}
    \begin{minipage}[c]{0.33\hsize}
    \includegraphics[clip,scale=0.9]{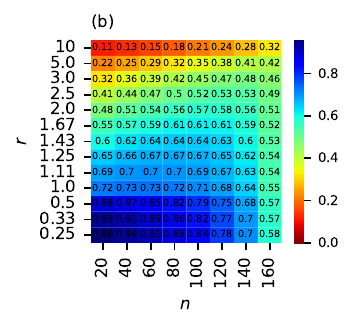}
    \end{minipage}
    \begin{minipage}[c]{0.33\hsize}
    \includegraphics[clip,scale=0.9]{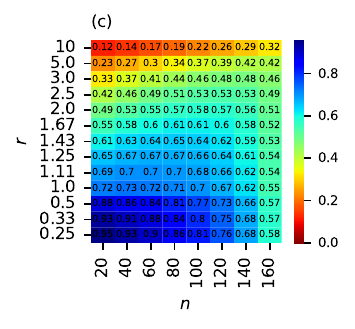}
    \end{minipage}\\
    \begin{minipage}[c]{0.33\hsize}
    \includegraphics[clip,scale=0.9]{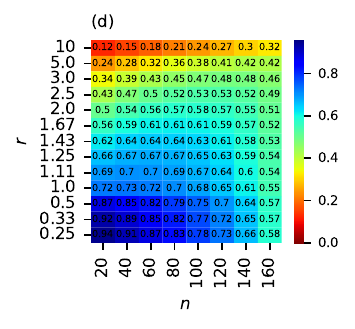}
    \end{minipage}
    \begin{minipage}[c]{0.33\hsize}
    \includegraphics[clip,scale=0.9]{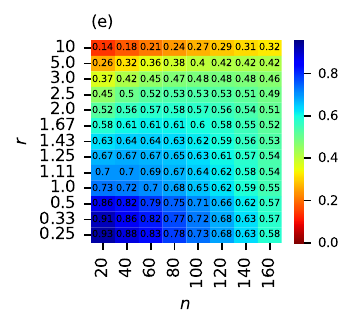}
    \end{minipage}
    \begin{minipage}[c]{0.33\hsize}
    \includegraphics[clip,scale=0.9]{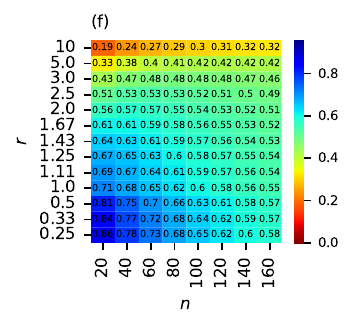}
    \end{minipage}
    \end{tabular}
    \caption{(Color online) Color map of the minimum energy gap $\Delta_\mathrm{min}$ distribution in $n$ versus $r$ space for the homogeneous fully connected ferromagnetic Ising model with $N=160$. The spacing of $r$ values is not necessarily uniform. $n=160$ represents the results without fixing spins. Each panel corresponds to a different error rate $p_\mathrm{err}$ for fixing spins: (a) $p_\mathrm{err}=0.0$, (b) $p_\mathrm{err}=0.1$, (c) $p_\mathrm{err}=0.2$, (d) $p_\mathrm{err}=0.3$, (e) $p_\mathrm{err}=0.4$, (f) $p_\mathrm{err}=0.5$.}
    \label{fig:error_res_n_energygap}
\end{figure*}
\begin{figure}[t]
    \centering
    \includegraphics[clip,scale=1.0]{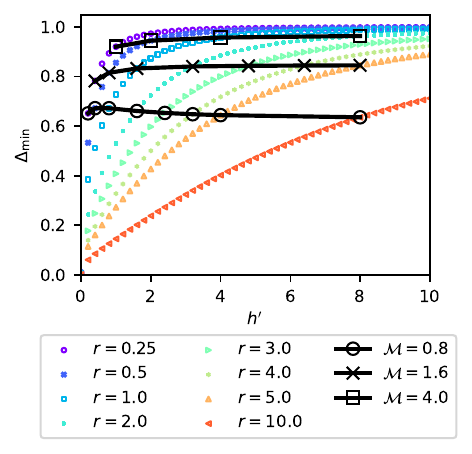}
    \caption{(Color online) Minimum energy gap $\Delta_\mathrm{min}$ at the thermodynamic limit for different $h^\prime$. Black lines represent $\Delta_\mathrm{min}$ for the same $h$ range condition with different $r$. The assumption is that spins are fixed up to the maximum value where the local magnetic fields reach $h/r$. This comparison considers the same value of $h/r$ across different numbers of spins but within the same parameter range. When $r$ is small, the number of fixed spins is small, while for large $r$, the number of fixed spins increases.}
    \label{fig:energy_gap_analysis_with_rescaling}
\end{figure}
\subsection{Energy gap analysis}
Since the minimum energy gap between the ground state and the first excited state, $\Delta_\mathrm{min}$, reflects the performance of quantum annealing near the adiabatic limit condition~\cite{Messiah1976adiabatictheorem, Tanaka-book}, we investigated its properties. 
Figure~\ref{fig:eg_rescaled} shows $\Delta_\mathrm{min}$ of the Hamiltonian after fixing spins with $p_\mathrm{err}=0$ and rescaling with $r$.
We perform exact diagonalization to obtain the minimum energy gap $\Delta_\mathrm{min}$.

The number of fixing spins $n$ at which $\Delta_\mathrm{min}$ is maximized varies for each $r$ as shown in Fig.~\ref{fig:eg_rescaled}.
Energy gap $\Delta_\mathrm{min}$ increases with $n$, consistent with a previous study~\cite{hattori2025advantages} without considering energy rescaling.
However, the fixing spins method does not necessarily lead to an expansion of $\Delta_\mathrm{min}$ when rescaling is considered.
For example, when $r=2.5$, the maximum $\Delta_\mathrm{min}$ appears at $n=140$.
These results indicate a trade-off between the energy gap expansion due to fixing spins and the reduction of the energy gap caused by rescaling.
% Additionally, for $r=10.0$,
Additionally, for $r=0.25$,
$\Delta_\mathrm{min}$ decreases as $n$ increases.
This demonstrates that the effects of fixing spins depend on the rescaling parameter.
The findings emphasize the importance of exploring the relationship between fixing spins and rescaling.

Next, we investigated the behaviour of $\Delta_\mathrm{min}$, including errors in fixing spins.
Figure~\ref{fig:error_res_n_energygap} presents a heatmap of $\Delta_\mathrm{min}$ for different values of $r$ and $n$. 
Each panel corresponds to a different $p_\mathrm{err}$.
As shown in Figure~\ref{fig:error_res_n_energygap}, as $p_\mathrm{err}$ increases or $r$ decreases, the optimal $n$ that maximizes $\Delta_\mathrm{min}$ becomes smaller. 
These results suggest that $r$ and $n$ must be carefully considered to achieve high-quality solutions.
For $r = 1.0$, fixing spins (lower $n$) improves $\Delta_\mathrm{min}$. 
In contrast, for $r = 10$, fixing spins reduces $\Delta_\mathrm{min}$. 
These findings highlight the necessity of considering the rescaling parameters when fixing spins and adjusting parameters accordingly.
Fixing spins has a positive effect at small $r$ but a negative effect at large $r$.

We further investigated the energy gap in the thermodynamic limit using the analytical method from the previous study~\cite{Seoane_2012} to assess the scalability of the numerical analysis.
The details of the analytical method are provided in appendix~\ref{appendix_eg}.
According to this method, the model is expressed as a harmonic oscillator. The energy gap derived from the method is given by:
\begin{align}
    \Delta(s)&=\left\{s^2\sin^2{\theta_0}\left(\frac{2h}{r}+\frac{1+\sin^2{\theta_0}}{s\sin{\theta}}\right)^2-\frac{(1+\sin^2{\theta_0})^2}{s^2\sin^2{\theta}}\right. \nonumber\\&
    +\left. s^2\sin^4{\theta_0}\frac{4J^2} {r^2}+s[2\sin^2{\theta_0}-2(1-s)\cos^3{\theta_0}]\frac{2J}{r}\right. \nonumber\\&
    \left. +4(1-s)^2\cos^2{\theta_0}\right\}^{\frac{1}{2}},\quad 0\leq s \leq 1,
    \label{eq:result_bozonize_analysis}
\end{align}
where $\theta_0$ is the rotation angle around the $y$-axis shown in Eq.~\eqref{eq:y_axis_rotation_composite}.
The angle $\theta_0$, which minimizes the energy, is numerically calculated.

Figure~\ref{fig:energy_gap_analysis_with_rescaling} shows the magnetic process of $\Delta_\mathrm{min}$ in the thermodynamic limit, where the minimum energy gap with respect to $s$ is denoted as $\Delta_\mathrm{min}$ in Eq.~\eqref{eq:result_bozonize_analysis}.
The fixing spins method introduces local magnetic fields without reducing the system size in the thermodynamic limit because the finite number of fixed spins is negligible compared to the total number of spins in the system.
Hence, increasing the local magnetic fields $h^\prime$ corresponds to increasing the number of fixed spins, which is related to a decrease in $n$ in Eq.~\eqref{eq:h_analysis_error}.
The error rate in fixing spins is also reflected by the reduction in local magnetic fields, as described by Eq.~\eqref{eq:h_analysis_error}.

In Fig.~\ref{fig:energy_gap_analysis_with_rescaling}, $\Delta_\mathrm{min}$ increases as $r$ decreases, in line with the increase in the local magnetic fields $h^\prime$ after fixing spins.
This indicates that smaller values of $r$ are more favourable for quantum annealing, assuming no restrictions on the parameter range.
The descending order of $h^\prime$ corresponds directly to the descending order of $\Delta_\mathrm{min}$.
However, it is important to note that the allowed parameter range differs for each $r$ in Fig.~\ref{fig:energy_gap_analysis_with_rescaling}, even though the parameter range is constrained in actual hardware.
Therefore, the comparison is more advantageous for settings with smaller $r$.

We also investigated $\Delta_\mathrm{min}$ while considering the restrictions of the parameter range.
The black lines in Fig.~\ref{fig:energy_gap_analysis_with_rescaling} illustrate a comparison of $\Delta_\mathrm{min}$ within the same $h$ range $[\mathcal{M}_\mathrm{min},\mathcal{M}_\mathrm{max}]$, where we assume $[-\mathcal{M},\mathcal{M}]$ with $|\mathcal{M}_\mathrm{min}|=|\mathcal{M}_\mathrm{max}|\equiv\mathcal{M}$.
We focus solely on the $h$ range because, in this numerical simulation, the $h$ range is always broader than the $J$ range after fixing spins.
According to Fig.~\ref{fig:energy_gap_analysis_with_rescaling}, the point at which the maximum energy gap occurs in the line of $\mathcal{M}=0.8$ is at $r=1.0$.
On the other hand, for the line of  $\mathcal{M}=4.0$, the point where the maximum energy gap occurs is $r=2.0$.
These results suggest that optimal rescaling parameters exist for different $h$ ranges.
Therefore, the effects of fixing spins in the thermodynamic limit also depend on the rescaling parameter.

When the $h$ range is wide, the fixing spins method effectively expands $\Delta_\mathrm{min}$.
In contrast, when the $h$ range is narrow, the fixing spins method becomes less effective because the rescaling effects, which reduce $\Delta_\mathrm{min}$, dominate over the benefits of expanding $\Delta_\mathrm{min}$ through fixing spins.
These findings are also supported by the finite-size Ising model results in Fig.~\ref{fig:error_res_n_energygap}.
The same considerations apply to the energy gap at finite sizes, even if the problem scale increases.
Therefore, the properties of the energy gap with fixing spins and rescaling are extensive with respect to size.

\section{\label{conclusion}Conclusion}
We investigated the effects of rescaling and fixing spins through experimental hardware and numerical simulations.
The experiment clarifies the impact of fixing spins and rescaling on a homogeneous fully connected ferromagnetic Ising model.
Our results demonstrate that fixing spins effectively leads to a low-energy state in quantum annealing hardware.
Additionally, fixing spins preserves the qubit chain generated by the minor embedding, which is an inherent feature when embedding the Ising model into hardware.

These findings suggest that the size-reduction method requires a broader range of parameters to address larger-scale problems on hardware.
Furthermore, to solve large-scale problems effectively, a fixing spins method with a low error probability is essential.
However, excessive reduction in problem size can adversely affect the energy scale, potentially compromising solution quality or hardware compatibility.

Based on these results, we can make proposals for hardware implementation.
The results emphasize that increasing the number of couplers for each qubit is a powerful strategy to prevent chain break.
Increasing the number of couplers shortens the chain length.
When the chain length is reduced, the local magnetic fields at each qubit become stronger as the distribution of the local magnetic fields across the qubits in a qubit chain is suppressed.
For the same range of magnetic fields and interactions in the Ising model, the ratio of the local magnetic fields to chain strength becomes larger, making chain breaks less likely. 
Therefore, increasing the number of couplers helps prevent chain breaks.

Energy gap analysis provided insights into the adiabatic limit, as the energy gap reflects the properties of this limit.
According to the numerical simulations, rescaling with a small parameter is not an effective way to increase the number of spins to fix.
From the perspective of $\Delta_\mathrm{min}$, a trade-off exists between the positive effects of fixing spins and rescaling.
The insight that a wider parameter range is necessary to overcome these issues aligns with the experimental results.

At the thermodynamic limit, this trade-off is also confirmed.
In situations with the same parameter range, a larger energy gap is obtained as the number of fixing spins increases.
These results are consistent with those observed in finite-size settings, indicating similar trends in larger-scale finite-size problems.
The trade-off insight suggests that expanding both the parameter range and the number of couplers is as crucial as increasing the number of qubits.

This study investigated the simplest setup to explore the complex relationships between fixing spins and rescaling.
Future work will investigate other models, such as the random Ising model and combinatorial optimization with constraints, which exhibit different properties of local magnetic fields that increase after fixing spins.

\begin{acknowledgment}
S.~T. was supported in part JSPS KAKENHI (Grant Number JP21K03391) and JST Grant Number JPMJPF2221, the Council for Science, Technology, and Innovation (CSTI) through the Cross-ministerial Strategic Innovation Promotion Program (SIP), ``Promoting the application of advanced quantum technology platforms to social issues'' (Funding agency: QST).
Human Biology-Microbiome-Quantum Research Center (Bio2Q) is supported by World Premier International Research Center Initiative (WPI), MEXT, Japan.
\end{acknowledgment}

\appendix
\section{\label{appendix_eg} Energy-gap analysis by bosonization in thermodynamic limit}
The total Hamiltonian with rescaling is written as follows:
\begin{align}
    \mathcal{H} (s) &= (1-s)\mathcal{H}_\mathrm{d} + s\mathcal{H}_\mathrm{p} \nonumber\\
    &=-(1-s)\sum_{i=1}^N\sigma_i^x + s\left(-\frac{h}{r} \sum_{i=1}^{N}\sigma_i^z -\frac{J}{rN}\sum_{1\leq i<j\leq N}\sigma_i^z\sigma_j^z\right)\nonumber\\
    &=-(1-s)2S_x  -s\frac{h}{r}2S_z -\frac{sJ}{2rN} \left(2S_z\right)^2+\mathrm{const},
    \label{eq:eg_analyze_Hp}
\end{align}
where $S_x$ and $S_z$ are $x$-component and $z$-component of the total spin variable in the composite system, respectively. 
Considering a rotation around the $y$-axis by an angle $\theta_0$, the composite spins can be expressed as follows:
\begin{align}
    \left(
    \begin{array}{r}
    S_x \\
    S_y \\
    S_z
    \end{array}
    \right)
    =\left(\begin{array}{rrr}
    -\sin{\theta_0} & 0 &\cos{\theta_0} \\
    0 & 1 & 0\\
    \cos{\theta_0}&0&\sin{\theta_0}
    \end{array}\right)
    \left(
    \begin{array}{r}
    \tilde{S}_x \\
    \tilde{S}_y \\
    \tilde{S}_z
    \end{array}
    \right).
    \label{eq:y_axis_rotation_composite}    
\end{align}
Subsequently, the total Hamiltonian after $y$-axis rotation is written as follows:
\begin{align}
    \mathcal{H} (s) &= -2(1-s)(-\sin{\theta_0}\tilde{S}_x+\cos{\theta_0}\tilde{S}_z)  \nonumber\\&-s\frac{2h}{r}(\cos{\theta_0}\tilde{S}_x+\sin{\theta_0}\tilde{S}_z) -\frac{2sJ}{rN} (\cos{\theta_0}\tilde{S}_x+\sin{\theta_0}\tilde{S}_z)^2 .   
    \label{eq:eg_H_analysis}
\end{align}
After applying the Holstein--Primacoff transformation\cite{HPtrans} to Eq.~\eqref{eq:eg_H_analysis}, the total Hamiltonian is expressed in terms of the creation and annihilation operators $a,a^{\dag}$.
The Holstein--Primacoff transformation is given by: 
\begin{align}
    \tilde{S}_z &=\frac{N}{2}-a^{\dag}a,\\
    \tilde{S}_+ &= \sqrt{N-a^{\dag}a}a=\sqrt{N}\sqrt{1-\frac{a^{\dag}a}{N}}a,\\
    \tilde{S}_- &= a^{\dag}\sqrt{N-a^{\dag}a}=a^{\dag}\sqrt{N}\sqrt{1-\frac{a^{\dag}a}{N}},\\
    \tilde{S}_x 
    &= \tilde{S}_{+}+\tilde{S}_{-}\nonumber\\
    &=\frac{\sqrt{N}}{2}\left(\sqrt{1-\frac{a^\dag a}{N}}a+a^\dag \sqrt{1-\frac{a^\dag a}{N}}\right).
    \label{eq:Holstein-Primacoff}
\end{align}
Moreover, we assume the thermodynamic limit $N\gg \langle a^{\dag}a\rangle$.
Therefore, the spin operator is represented as follows:
\begin{align}
    \tilde{S}_z &=\frac{N}{2}-a^{\dag}a,\\
    \tilde{S}_x &\approx \frac{\sqrt{N}}{2} (a+a^{\dag}).
    \label{eq:Holstein-Primacoff_appro}
\end{align}
By expanding the Eq.~\eqref{eq:eg_analyze_Hp} with spin operator after $y$-axis rotation, the total Hamiltonian is written as follows:
\begin{align}
    \mathcal{H} (s)=&\left[2(1-s)\cos{\theta_0}-s\frac{2h}{r}\sin{\theta_0}-\frac{sJ\sin^2{\theta_0}}r\right]\frac{N}{2}\nonumber\\ 
    &+\left[2(1-s)\sin{\theta_0}-s\frac{2h}{r}\cos{\theta_0}+\right. \nonumber \\
    &-\left.s\frac{2J}{r}\cos{\theta_0}\sin{\theta_0}\right](a+a^{\dag})\frac{\sqrt{N}}{2}+2(1-s)\cos\theta_0\nonumber\\  
    &+s\frac{2h}{r}\sin{\theta_0}-s\frac{J}{r}\cos^2\theta_0+s\frac{2J}{r}\sin^2{\theta_0}a^\dag a\nonumber\\
    &-s\frac{J}{2r}\cos^2\theta_0\left[(a^\dag)^2+a^2+1\right]\nonumber\\
    &-\frac{sJ}{r\sqrt{N}}\cos{\theta_0}\sin{\theta_0}[a^{\dag}a(a+a^{\dag})+(a^\dag+a)a^\dag a] \nonumber\\
    &-s\frac{2J}{2rN}\sin^2{\theta_0} (a^{\dag}a)^2.
    \label{eq:Holstein-Primacoff_N_order}
\end{align}
Here, we assume that $\frac{1}{\sqrt{N}}\rightarrow 0$. 
Transforming Eq.~\eqref{eq:Holstein-Primacoff_N_order} using the commutation relation $aa^{\dag}-a^{\dag}a=1$ yields:
\begin{align}
    \mathcal{H} (s)\approx&\left[2(1-s)\cos{\theta_0}-s\frac{2h}{r}\sin{\theta_0}-\frac{sJ\sin^2{\theta_0}}{r}\right]\frac{N}{2}\nonumber\\ 
    &+\left[2(1-s)\sin{\theta_0}-s\frac{2h}{r}\cos{\theta_0}+\right. \nonumber \\
    &-\left.s\frac{2J}{r}\cos{\theta_0}\sin{\theta_0}\right](a+a^{\dag})\frac{\sqrt{N}}{2}+2(1-s)\cos\theta_0\nonumber\\ 
    &+s\frac{2h}{r}\sin{\theta_0}-s\frac{J}{r}\cos^2\theta_0+s\frac{2J}{r}\sin^2{\theta_0}a^\dag a\nonumber\\
    &-s\frac{J}{2r}\cos^2\theta_0\left[(a^\dag)^2+a^2+1\right].
    \label{eq:Holstein-Primacoff_N_order_appro}
\end{align}
Here, we define $e, \delta$ and $\gamma$ as  
\begin{align}
    \label{eq:e}
    e \equiv& \frac{1}{2}\left[-2(1-s)\cos{\theta_0}-s\frac{2h}{r}\sin{\theta_0}-s\frac{J}{r}\sin^2{\theta_0}\right],\\
    \label{eq:delta}
    \delta \equiv& 2(1-s)\cos{\theta_0}+s\frac{2h}{r}\sin{\theta_0}+s\frac{2J}{r}\sin^2{\theta_0}-s\frac{J}{r}\cos^2{\theta_0},\\
    \label{eq:gamma}
    \gamma \equiv&-\frac{sJ}{2r}\cos^2{\theta_0}.
\end{align}
We set $\theta_0$ as $e$ is the minimum value.
\begin{align}
    \label{eq:minimum_theta_0}
    \frac{\partial e}{\partial \theta_0}=\frac{1}{2}\left[2(1-s)\sin{\theta_0}-s\frac{2h}{r}\cos{\theta_0}-s\frac{2J}{r}\cos{\theta_0}\sin{\theta_0}\right]= 0.
\end{align}
Since calculating $\theta_0$ that minimizes $e$ analytically is challenging, we solve for $\theta_0$ numerically.
Following Eq.~\eqref{eq:e}, Eq.~\eqref{eq:delta}, and Eq.~\eqref{eq:gamma}, the total Hamiltonian is expressed as follows:
\begin{align}
    \mathcal{H} (\delta,\gamma) \approx eN +\gamma +\gamma[a^2+(a^{\dag})^2]+\delta a^{\dag}a.
\end{align}
Lastly, the Bogoliubov transformation\cite{bogoliubov} is performed to $a,a^{\dag}$.
\begin{align}
    &a = \cosh{\frac{\Theta}{2}}b+\sinh{\frac{\Theta}{2}}b^{\dag},\\
    &a^{\dag}=\cosh{\frac{\Theta}{2}}b^{\dag}+\sinh{\frac{\Theta}{2}}b,
\end{align}
where, $b,b^{\dag}$ is the new creation-annihilation operator after Bogoliubov transformation. 
Moreover, $\Theta$ satisfies the following equation:
\begin{align}
    \tanh{\Theta}\equiv -\frac{2\gamma}{\delta}\equiv\epsilon.
\end{align}
The total Hamiltonian after the Bogoliubov transformation is written as follows:
\begin{align}
\label{eq:harmonic}
    \mathcal{H} (\delta,\gamma) \approx eN +\gamma +\frac{\delta}{2} (\sqrt{1-\epsilon^2}-1)+\Delta b^{\dag}b.
\end{align}
Equation~\eqref{eq:harmonic} represents the Hamiltonian of the harmonic oscillator. 
Hence, the energy gap is expressed as follows:
\begin{align}
    \Delta(s)&=\delta\sqrt{1-\epsilon^2}\nonumber\\
    &=\left\{s^2\sin^2{\theta_0}\left(\frac{2h}{r}+\frac{1+\sin^2{\theta_0}}{s\sin{\theta_0}}\right)^2-\frac{(1+\sin^2{\theta_0})^2}{s^2\sin^2{\theta_0}}\right. \nonumber\\&
    +\left. s^2\sin^4{\theta_0}\frac{4J^2}{r^2}+s[2\sin^2{\theta_0}-2(1-s)\cos^3{\theta_0}]\frac{2J}{r}\right. \nonumber\\&
    \left. +4(1-s)^2\cos^2{\theta_0}\right\}^{\frac{1}{2}},\quad 0\leq s\leq 1.
    \label{eq:result_bozonize_analysis_appendix}
\end{align}

\section{Effective annealing schedule after rescaling\label{appendix_annealing_sche}}
\begin{figure}[t]
    \centering
    \includegraphics[clip,scale=1.0]{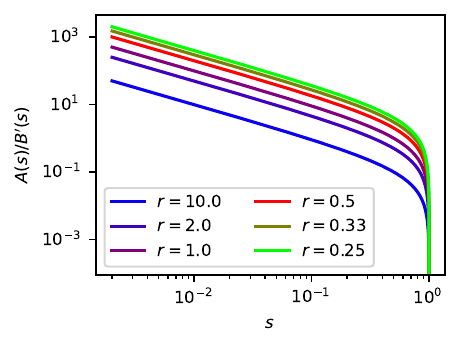}
    \caption{(Color online) Annealing schedule after fixing spins with rescaling. The vertical axis represents the ratio of the transverse local magnetic fields for the problem Hamiltonian. At $t=0$, the ratio becomes infinite.}
    \label{fig:rescaled_schedule}
\end{figure}

One effect of rescaling is the modification of the effective annealing schedule.
By dividing the Hamiltonian by a rescaling parameter $r$, the effective annealing schedule is altered, as shown in Eq.~\eqref{eq:rescaling_H}.
Consequently, the annealing schedule changes as follows:
\begin{align}
    \label{eq:effective_annealing_schedule}
    \mathcal{H}^\prime(s) =& A(s)\mathcal{H}^\prime_\mathrm{d}+\frac{B(s)}{r}\mathcal{H}^\prime_\mathrm{p},\\
    =&A(s)\mathcal{H}^\prime_\mathrm{d}+B^\prime(s)\mathcal{H}^\prime_\mathrm{p},
\end{align}
where $B^\prime(t)$ represents the effective annealing schedule after rescaling.
Eq.~\eqref{eq:effective_annealing_schedule} shows that rescaling alters only the gradient of the annealing schedule, implying that rescaling modifies the required annealing time.
When the rescaling parameter is small, the crossover point, where $A(s)/B(s)=1$, is delayed.
Figure \ref{fig:rescaled_schedule} illustrates the changes in the annealing schedule. 
Numerous studies\cite{Morita_2007} have highlighted the critical role of the annealing schedule in quantum annealing.
Specifically, the study~\cite{Morita_2007} suggests that the $k$~th derivative of the Hamiltonian at $t=0$ and $t=\tau$ determines the upper bound of the excitation probability from the ground state to the first excited state.
As $r$ increases, the speed of the annealing schedule remains unaffected, as illustrated in Fig.~\ref{fig:rescaled_schedule}.

\bibliography{71494}
\bibliographystyle{jpsj}

\end{document}